\documentclass[pra,superscriptaddress,amsmath,amssymb,12pt]{revtex4}
\usepackage{graphicx}
\usepackage{caption2}
\usepackage{mathrsfs,graphicx,color,float,indentfirst,textcomp}
\usepackage{latexsym,lscape,enumerate,hyperref}

\begin{document}
\title{One-way quantum deficit for $2\otimes d$ systems}
\author{Biao-Liang Ye}
\affiliation{School of Mathematical Sciences, Capital Normal University,
Beijing 100048, China}
\author{Shao-Ming Fei}
\affiliation{School of Mathematical Sciences, Capital Normal University,
Beijing 100048, China}
\affiliation{Max-Planck-Institute for Mathematics in the Sciences,
04103 Leipzig, Germany}

\begin{abstract}
We investigate one-way quantum deficit for $2\otimes d$ systems.
Analytical expressions of one-way quantum deficit under both
von Neumann measurement and weak measurement are presented.
As an illustration, qubit-qutrit systems are studied in detail.
It is shown that there exists non-zero one-way quantum deficit
even quantum entanglement vanishes.
Moreover, one-way quantum deficit via weak measurement
turns out to be weaker than that via von Neumann measurement.
The dynamics of entanglement and one-way quantum deficit under
dephasing channels is also investigated.
\end{abstract}

\maketitle

\section{Introduction}
Quantum entanglement is one of the most important quantum correlations and plays
a fundamental role in quantum information science \cite{Horodecki2009,Amico2008}.
Beyond entanglement, quantum discord \cite{Ollivier2001,Henderson2001}
plays a key role in some quantum speed-up for quantum
information tasks \cite{Merali2011}, for instance,
in assisted optimal state discrimination only one side discord is required, while the entanglement is not
necessary \cite{Roa2011,Li2012}.
Other different measures in quantifying quantum correlations \cite{Modi2012}, such as
one-way quantum deficit \cite{Oppenheim2002,Horodecki2005},
quantum dissonance \cite{Modi2010}, geometric discord \cite{Dakic2010},
measurement-induced  nonlocality \cite{Luo2011} have been also provided.
Nonetheless, usually it is formidably difficulty to get analytical results for these quantum correlations.
Analytical expressions  of quantum discord \cite{Li2011,Lu2011,Chen2011} seem to be extremely
hard due to the  optimization involved \cite{Huang2013,Huang2014}.
Only a few analytical results for the simplest two-qubit systems
have been worked out \cite{Luo2008a,Maldonado-Trapp2015,Yurishchev2014,Yurischev2014,Yurischev2015}.

Recently, in Ref. \cite{Chanda2015} the authors shew that analogous to quantum discord,
one-way quantum deficit exhibits also frozen phenomenon.
The one-way quantum deficit and quantum discord in $XX$ spin chains have been investigated in \cite{Ciliberti2013}.
And the explicit relationship between quantum discord and one-way
quantum deficit has been studied in \cite{Ye2016a}.
Similar to quantum discord, it is hard to derive analytical expressions of one-way quantum deficit of general
two-qubit systems. The upper bound of one-way quantum deficit
is shown to be the entropy of the measured subsystem \cite{Shao2013}. Partial analytical
expressions of one-way quantum deficit
of five-parameter two-qubit $X$ states have been provided in Ref \cite{ye2016}.

In this paper, we study the one-way quantum deficit for $2\otimes d$ (qubit-qudit) systems.
We provide the analytical results of one-way quantum
deficit for a two-parameter class of states in
$2\otimes d$ quantum systems with $d\ge 3$.
Moreover, we utilize the weak measurement \cite{Aharonov1988}  to
investigate the one-way quantum deficit for the systems. Generally
weak measurement exhibits amplifying roles \cite{Singh2014}. However, we find
the one-way quantum deficit via weak measurement
is weaker than that via von Neumann measurement.
We also study the decoherence of one-way quantum deficit via
von Neumann measurement and weak measurement for qubit-qutrit systems.

\section{One-way quantum deficit via von Neumann measurement}

One-way quantum deficit is related to extracting
work from a correlated system to a heat bath under
local operations \cite{Oppenheim2002}.
Consider Alice ($A$) and Bob ($B$) share a bipartite quantum system $\rho_{AB}\in \mathcal{H}^2\otimes \mathcal{H}^d$
in $2$ and $d$ dimensional spaces $\mathcal{H}^2$ and $\mathcal{H}^d$, respectively.
Let $\{P_i\}$ be local von Neumann (projective) measurement,
$P_i P_j=\delta_{ij}P_i$, $\sum_i P_i=I$, with $I$ the identity operator.
The one-way quantum deficit is defined as the minimal
increase of entropy after the projective measurement performing on the subsystem $A$ \cite{Streltsov2011},
\begin{equation}\label{def}
	\overset{\rightharpoonup}{\vphantom{a}\smash{\Delta}}(\rho_{AB})
	=\min_{\{P_j^A\}}S(\rho_{AB}')-S(\rho_{AB}),
\end{equation}
where
$\rho_{AB}'=\sum_j (P_j^A\otimes I) \rho_{AB}  (P_j^A\otimes I)$ is
the state after measurement on $A$,
$S(\rho)=-{\rm Tr}\rho\log_2\rho$ is
the von Neumann entropy of the state $\rho$,
and the minimum is taken over all possible projective measurements $\{P_j^A\}$.
The one-way information deficit is non-negative and
zero for classical-quantum correlated states.

A two-parameter family of states in $2\otimes d$ quantum systems was first
introduced in \cite{Chi2003},
\begin{eqnarray}\label{state}
\rho_{r,t}&&=r\sum_{i=0}^1\sum_{j=2}^{d-1}|ij\rangle\langle ij|
+s(|\phi^+\rangle\langle \phi^+|+|\phi^-\rangle\langle \phi^-|\nonumber\\
&&+|\psi^+\rangle\langle\psi^+|)+t|\psi^-\rangle\langle\psi^-|,
\end{eqnarray}
where $\{|ij\rangle: i=0, 1, j=2, 3, \dots, d-1\}$
are orthonormal bases for the $2\otimes d$ quantum systems
and the four Bell bases are given as follows
\begin{eqnarray}\nonumber
|\phi^\pm\rangle=\frac{1}{\sqrt{2}}(|00\rangle\pm|11\rangle),~~~
|\psi^\pm\rangle=\frac{1}{\sqrt{2}}(|01\rangle\pm|10\rangle).
\end{eqnarray}
The parameters satisfy $2(d-2)r+3s+t=1$ with
$0\le r\le 1/(2d-4)$.
It has been proven that any $2\otimes d$ states can be transformed
into $\rho_{r,t}$ with the help of local operations
and classical communication (LOCC) \cite{Chi2003}.
The quantum discord for such states have been studied in Refs. \cite{Ali2010,Vinjanampathy2012}.

Now, let us turn to calculate one-way quantum deficit
for the state (\ref{state}).
We perform measurements on subsystem $A$ by projective operators
$P_k^A=|k'\rangle\langle k'|$, $k\in\{0, 1\}$,  where
\begin{equation}\label{pm}
\begin{array}{l}
|0'\rangle=\cos(\theta/2)|0\rangle-e^{-i\phi}\sin(\theta/2)|1\rangle,\\[2mm]
|1'\rangle=e^{i\phi}\sin(\theta/2)|0\rangle+\cos(\theta/2)|1\rangle.
\end{array}
\end{equation}
The projective measurement bases are described by the
angles $\theta$ and $\phi$ of the Bloch sphere,
with $\theta\in[0,\pi]$ and $\phi\in[0,2\pi)$.

To treat one-way quantum deficit, the key problem is to
minimize the first term in (\ref{def}).
However, the eigenvalues of
$\sum_j (P_j^A\otimes I) \rho_{r,t}  (P_j^A\otimes I)$
do not contain the measurement parameters $\theta$ and $\phi$.
That is to say, for $2\times d$ systems, the one-way
quantum deficit is independent of projective
measurement and we do not need to do the minimization.
Namely, the result is optimal under any projective measurement.
The eigenvalues of the post measured state are given
by $\{s,s, \frac{s+t}{2}, \frac{s+t}{2}, r, r, \cdots, r\}$.
Taking into account that $S(\rho_{r,t})=-[3s\log_2s+t\log_2t+2(d-2)r\log_2r]$,
we obtain the analytical expressions of one-way quantum deficit
for $2\otimes d$ states,
\begin{equation}
\overset{\rightharpoonup}{\vphantom{a}\smash{\Delta}}
=s\log_22s+t\log_22t-(s+t)\log_2(s+t).
\end{equation}
It turns out that one-way quantum deficit
for the $2\otimes d$ systems is same as the quantum discord of
the two-parameter states \cite{Ali2010,Vinjanampathy2012}.

\section{One-way quantum deficit via weak measurement}
Weak measurement was formulated
in Ref. \cite{Aharonov1988} by using the pre and
post-selected quantum systems. In Ref. \cite{Oreshkov2005}
the authors constructed weak measurement operators,
\begin{eqnarray}\label{wm}\nonumber
	q(+x)=\sqrt{\frac{1-\tanh[x]}{2}}M_0+\sqrt{\frac{1+\tanh[x]}{2}}M_1,\\\nonumber
	q(-x)=\sqrt{\frac{1+\tanh[x]}{2}}M_0+\sqrt{\frac{1-\tanh[x]}{2}}M_1,
\end{eqnarray}
where $x$ is a parameter describing the strength of
the measurement, $M_0$ and $M_1$ are the two orthogonal
projectors satisfying $M_0+M_1=I$ and $q(+x)^\dagger q(+x)+q(-x)^\dagger q(-x)=I$.
Much attention has been paid to weak measurement both
theoretically and experimentally \cite{Dressel2014}.

Now we study one-way quantum deficit under weak measurement.
Instead of projective measurement, under weak measurement the post measured
state has the form
$\rho'_{r,t}=\sum_{+x,-x}[q(x)\otimes I]\cdot\rho_{r,t}\cdot [q(x)\otimes I]^\dagger$.
The eigenvalues of this state is given by
$\{\frac12(s+t+(s-t){\rm sech}[x],\frac12(s+t-(s-t){\rm sech}[x], s, s, r, r, \cdots, r\}$.
Thus, the one-way quantum deficit via weak measurement is given by
\begin{eqnarray}
\overset{\rightharpoonup}{\vphantom{a}\smash{\Delta}}_w
	&&=-\sum_{i=0,1}\Lambda_i\log_2\Lambda_i+s\log_2s+t\log_2t,
\end{eqnarray}
where $\Lambda_i=\frac12[s+t+(-1)^i(s-t){\rm sech}[x]].$

Now, we have derived the analytical formulae of one-way quantum deficit
under projective measurement and weak measurement, respectively.
It is observed that the analytical expressions of
one-way quantum deficit under weak measurement
or projective measurement are independent of the dimension $d$.
In the following we investigate the relationship between quantum entanglement
and one-way quantum deficit, as well as their evolution under noisy channels.

\section{Entanglement, one-way quantum deficit in qubit-qutrit systems}

We consider qubit-qutrit systems ($d=3$). The qubit-qutrit states are given by
\begin{eqnarray}
	 \sigma_{r,t}&&=r(|02\rangle\langle02|+|12\rangle\langle12|)
	 +s(|\phi^+\rangle\langle\phi^+|+|\phi^-\rangle\langle\phi^-|\nonumber\\
	 &&+|\psi^+\rangle\langle\psi^+|)+t|\psi^-\rangle\langle\psi^-|.
\end{eqnarray}
The geometric discord of such states under various noise channels
have been studied in Ref. \cite{Wei2013}.

For $2\otimes 3$ systems the positive partial
transposition (PPT) criterion is the necessary and sufficient condition for
separability \cite{Peres1996,Horodecki1996}.
We use the \emph{negativity} $N$ as the measure of entanglement \cite{Chi2003},
\begin{equation}
\emph{N}(\sigma)=\max\{0,\|\sigma^{T_B}\|_1	-1\},
\end{equation}
where ${T_B}$ stands for the partial transposition with respect to the
subsystem $B$, and $\sigma^{T_B}$ is the partial transposition state of $\sigma$,
$\|\sigma\|_1={\rm Tr}[\sqrt{\sigma^\dagger\sigma}]$
denotes the trace norm of $\sigma$.
For the qubit-qutrit state $\sigma_{r,t}$ the negativity
is given by $N(\sigma_{r,t})=\max\{0, 2(r+t)-1\}$.

\begin{figure}[h]
  \includegraphics[width=8.4cm]{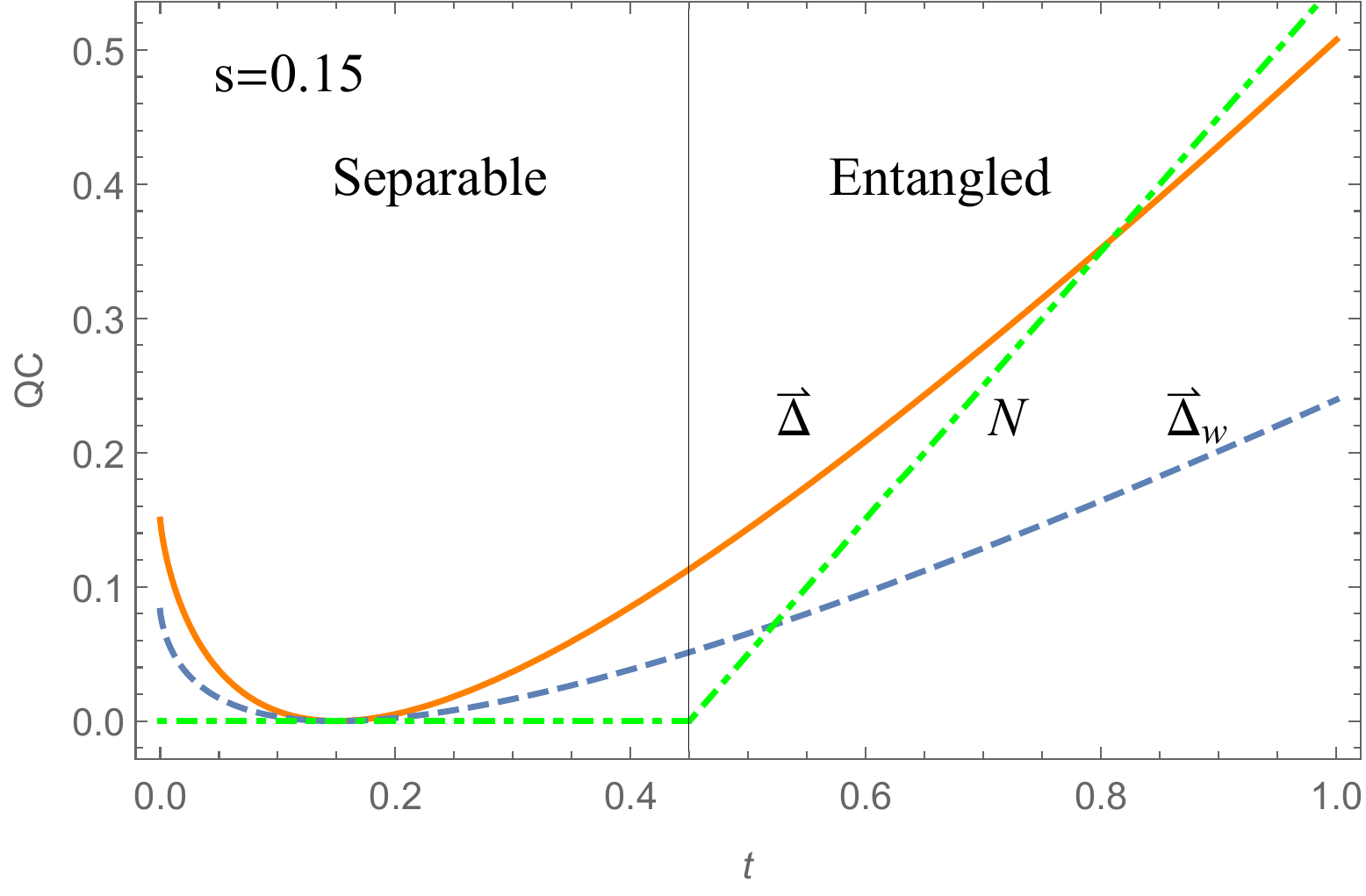}\\
  \caption{(Color online) Solid orange line for one-way quantum deficit via projective measurement
  $\overset{\rightharpoonup}{{\Delta}}$, dashed blue line for
  weak quantum deficit $\overset{\rightharpoonup}{{\Delta}}_w$, and
  dotted-dashed green line for negativity {\it N}.}\label{cor}
\end{figure}

Take $s=0.15$. The relationship between the negativity
and one-way quantum deficit is shown in Fig.\ref{cor}.
The state is separable for $t\leq 0.45$ and entangled for $t>0.45$.
For separable states, the one-way quantum deficit via projective and weak measurements
could be still greater than zero.
The weak quantum deficit (dashed blue line) is weaker than
one-way quantum deficit via von Neumann measurement (solid orange line).

Now we consider decoherence of qubit-qutrit systems under dephasing channels.
After the dephasing channels the qubit-qutrit state $\sigma_{r,t}$ is transformed to be
\begin{eqnarray}\nonumber
\sigma_{r,t}'=\sum_{i=0}^1\sum_{j=0}^2E_i\otimes F_j \cdot\sigma_{r,t} \cdot E_i^\dagger\otimes F_j^\dagger,
\end{eqnarray}
where
\begin{eqnarray}\nonumber
E_0=\left(
\begin{array}{cc}
 1 & 0 \\
 0 & \sqrt{1-\gamma _A} \\
\end{array}
\right),~~~
E_1=\left(
\begin{array}{cc}
 0 & 0 \\
 0 & \sqrt{\gamma _A} \\
\end{array}
\right),
\end{eqnarray}
and
\begin{eqnarray}\nonumber
&&F_0=\left(
\begin{array}{ccc}
 1 & 0 & 0 \\
 0 & \sqrt{1-\gamma _B} & 0 \\
 0 & 0 & \sqrt{1-\gamma _B} \\
\end{array}
\right),~~~
F_1=\left(
\begin{array}{ccc}
 0 & 0 & 0 \\
 0 & \sqrt{\gamma _B} & 0 \\
 0 & 0 & 0 \\
\end{array}
\right),\nonumber\\[2mm]
&&F_2=\left(
\begin{array}{ccc}
 0 & 0 & 0 \\
 0 & 0 & 0 \\
 0 & 0 & \sqrt{\gamma _B} \\
\end{array}\nonumber
\right).
\end{eqnarray}
The parameters $\gamma_A=1-e^{-\tau\Gamma_A}$ and
$\gamma_B=1-e^{-\tau\Gamma_B}$, with $\Gamma_{A(B)}$
the decay rate of the subsystem $A(B)$ and $\gamma_{A(B)}\in [0,1]$.

The one-way quantum deficit of state $\sigma_{r,t}'$ can be calculated directly
from the optimal projective measurement (\ref{pm}) with $\theta=0$ and arbitrary $\phi$,
which is given by
\begin{eqnarray}
\overset{\rightharpoonup}{\vphantom{a}\smash{\Delta}}(\sigma_{r,t}')=
\sum_{j=0}^1\lambda_j\log_2\lambda_j-(s+t)\log_2\frac12(s+t).
\end{eqnarray}
where
$$\lambda_j=\frac12[s+t+(-1)^j(s-t)\sqrt{(1-\gamma_A)(1-\gamma_B)}].$$




Similarly, under this decoherence channel the one-way quantum deficit
via weak measurement is given by
\begin{eqnarray}
\overset{\rightharpoonup}{\vphantom{a}\smash{\Delta}}_w(\sigma_{r,t}')
	=\sum_{j=0}^1[\eta_j \log_2\eta_j-\xi_j \log_2 \xi_j],
\end{eqnarray}
where $$\eta_j=\frac12[s+t+(-1)^j(s-t)\sqrt{(1-\gamma_A)(1-\gamma_B)}],$$
and $$\xi_j=\frac12[s+t+(-1)^j(s-t){\rm sech}[x]\sqrt{(1-\gamma_A)(1-\gamma_B)}].$$

The negativity of $\sigma_{r,t}'$ has the form
\begin{equation}
\emph{N}(\sigma_{r,t}')=\max\left\{0, \frac13[2(2r+t-1)+(2r+4t-1)\sqrt{(1-\gamma_A)(1-\gamma_B)}]\right\}.	
\end{equation}

\begin{figure}[h]
  \includegraphics[width=8.4cm]{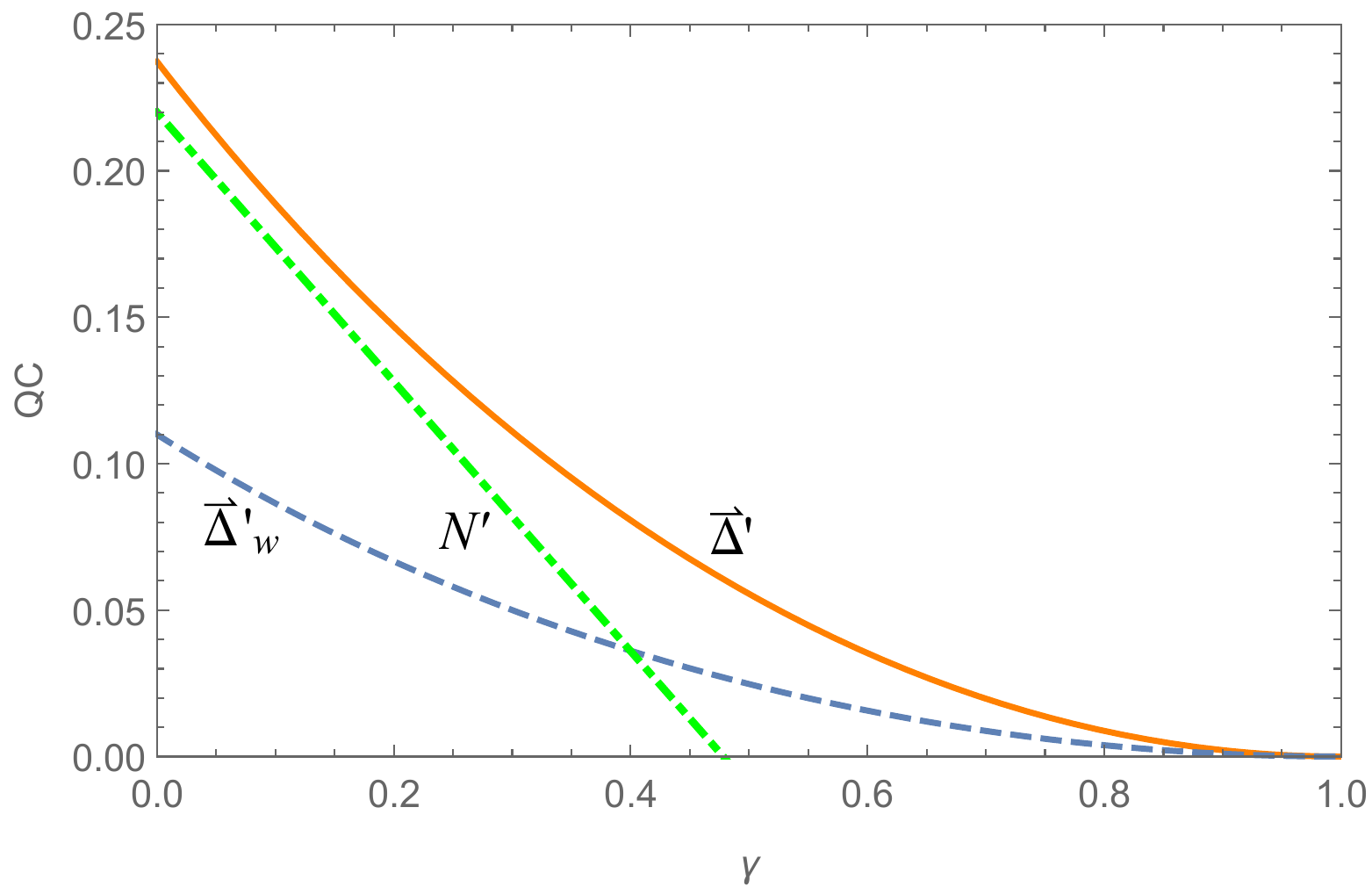}\\
  \caption{(Color online) Under dephasing channels, the quantum
  correlation for qubit-qutrit systems.
  The decoherence of one-way quantum deficit via projective measurement
  $\overset{\rightharpoonup}{{\Delta'}}$ is depicted by solid orange line.
  Weak quantum deficit $\overset{\rightharpoonup}{{\Delta'}}_w$
  is described in dashed blue line. The dotted-dashed green line
  denotes negativity {$N'$}.
  We suppose $\gamma_A=\gamma_B=\gamma$ and set
  $r=0.03, s=0.12, t=0.58, x=0.8$. }\label{dym}
\end{figure}

The dynamics of the system under dephasing channel can be seen in Fig. \ref{dym}.
In finite time, entanglement sudden death
happens (dotted-dashed green line), while one-way
quantum deficit under projective or weak measurements vanish gradually.
Moreover, under the whole decoherence dynamical process the weak quantum deficit
is also always weaker than the one-way quantum deficit via projective
measurement.

\section{Conclusions}
We have extended previous studies on one-way quantum deficit for two-qubit systems to
the case of $2\otimes d$ systems.
We have provided analytical expressions of one-way quantum
deficit under both projective measurement and weak measurement.
It has been shown that there still exits non-zero one-way quantum deficit for separable states.
In particular, we have investigated the quantum entanglement (negativity)
and quantum deficits for qubit-qutrit systems.
It has been found that the one-way quantum deficit via weak measurement is weaker
than the one under projective measurement.
Under the decoherence of dephasing channel, one sees the entanglement sudden death,
while one-way quantum deficits do not vanish suddenly.
Our results could help to understand the one-way quantum deficit.
Such approach may be also used to investigate quantum correlations for multipartite systems.

\section*{Acknowledgment}
This work is supported by NSFC under number 11275131.

\end{document}